\documentstyle[pra,aps,psfig]{revtex}
\newcommand{\beq}{\begin{equation}}
\newcommand{\eeq}{\end{equation}}
\newcommand{\beqa}{\begin{eqnarray}}
\newcommand{\eeqa}{\end{eqnarray}}
\newcommand{\ba}{\begin{array}}
\newcommand{\ea}{\end{array}}

\begin{document}
\draft

\twocolumn[\hsize\textwidth\columnwidth\hsize\csname
@twocolumnfalse\endcsname

\widetext 

\title{Pulsed Macroscopic Quantum Tunneling \\
of Falling Bose Condensates} 
\author{L. Salasnich$^{1}$, A. Parola$^{2}$ and L. Reatto$^{1}$} 
\address{$^{1}$Istituto Nazionale per la Fisica della Materia, 
Unit\`a di Milano Universit\`a, \\ 
Dipartimento di Fisica, Universit\`a di Milano, \\
Via Celoria 16, 20133 Milano, Italy \\ 
$^{2}$Istituto Nazionale per la Fisica della Materia, Unit\`a di Como, \\ 
Dipartimento di Scienze Fisiche, Universit\`a dell'Insubria, \\
Via Lucini 3, 23100 Como, Italy} 

\maketitle

\begin{abstract} 
We investigate macroscopic quantum tunneling of a Bose 
condensate and how it is affected by the interatomic 
interaction. We study the dynamics of a condensate 
falling under gravity and scattering on a Gaussian potential barrier 
that models a mirror formed by a far-detuned sheet of light. 
We observe bouncing, interference and quantum tunneling 
of the condensate. We find that the tunneling rate is 
very sensitive to the interatomic interaction 
and to the shape of the condensate. Under many conditions 
the tunneling rate is strongly enhanced by 
the interaction as achieved, for instance, by increasing 
the number of condensed particles. In a quasi 1D situation 
the tunneling pulse displays two peaks. 
The quantum tunneling can be quasi-periodic and in this way 
one could generate coherent Bose condensed atomic pulses. 
\end{abstract}

\vskip 0.5cm

]

\narrowtext

\vskip 0.5cm 

\section{Introduction} 

Bose condensates, as achieved in trapped clouds of 
alkali-metal atoms [1-3], are ideal systems to study quantum 
phenomena at macroscopic level [4], 
like interference [5] and diffraction of matter waves. 
Tunneling is one of the striking quantum 
phenomena, which is important also in many applications. 
In the case of non-interacting particles at very low temperatures, 
when almost all particles are in the condensate, the tunneling 
of the condensate has the same character as in the case of 
a single particle. 
In this paper we address the question of how tunneling is affected 
by the interatomic interaction. The case of a condensate 
in a double-well potential has been already theoretically 
investigated and it is known that the interatomic 
interaction tends to suppress tunneling [6-10]. 
This effect can be so strong that under certain conditions 
it has been predicted that tunneling is suppressed completely 
and the condensate remains self-trapped in one of the minima 
of the potential. Here we show that under other conditions the tunneling 
rate can be {\it enhanced} by the interatomic interaction. 
\par 
The case we study is that of a condensate impinging 
on a potential barrier that could model a mirror 
formed by a far-detuned sheet of light. 
We find that the interatomic interaction and the 
geometrical aspects of the system, like the aspect ratio 
of the cloud or the fact the the cloud remains trapped 
in the transverse directions, have a strong effect on the 
tunneling probability. This can be either enhanced or 
depressed and this effect can be very strong when the number 
of particles is large. The shape of the transmitted pulse 
can differ strongly from the initial shape and 
under certain conditions it gets a double peak structure. 
\par 
The specific system we consider is a condensate falling under gravity 
and scattering on a Gaussian potential barrier. 
In a recent experiment, it has been shown the bouncing of condensates 
on the mirror formed by a far-detuned sheet of light
when the strength of the energy barrier 
is much larger than the kinetic energy 
of the condensate [11]. 
Here we reproduce the bouncing effect 
and also predict that, 
when the energy barrier becomes comparable to the kinetic 
energy, macroscopic quantum tunneling can be observed. 
We consider the case in which the trapping potential in the 
horizontal directions remains active during the fall as well as 
the case in which the trapping potential is turned off at an 
initial time in all directions. We make comparison also with 
the behavior of a one dimensional case. 

\section{Numerical procedure} 

The zero-temperature dynamics of a low-density 
Bose condensate can be accurately described by the time-dependent 
Gross-Pitaevskii equation (TDGPE) [12,13] 
for the macroscopic wavefunction (order parameter) 
$\psi ({\bf r},t)$ of the condensate in an external potential 
$V_{ext}({\bf r})$. When the wavefunction is normalized to unity then 
the parameter of the nonlinear term of the TDGPE is 
$\gamma={4\pi \hbar^2 a_s(N-1)/m}$, 
where $a_s$ the s-wave scattering length and $N$ is the number 
of condensed atoms.  
\par 
In our problem, a Bose condensate is initially trapped 
in a harmonic potential and is under the action of gravity. 
In addition there is a Gaussian barrier centered at $z=0$. 
Therefore at $t\leq 0$ the external potential reads 
\beq 
V_{ext}^{<}(\rho,z)={m\over 2} (\omega_{\rho}^2\rho^2 
+\omega_z^2(z-z_0)^2) + m g z + U e^{-{z^2\over \sigma^2}} \; , 
\eeq 
where $\rho=(x^2+y^2)^{1/2}$ and $z$ are the cylindrical 
coordinates, $\omega_{\rho}$ and $\omega_z$ are the frequencies 
of the trapping harmonic potential, $z_0$ is the position 
of the trap minimum along the $z$ axis, 
$g$ is the gravity acceleration, $U$ is the height of 
the potential barrier and $\sigma$ its width. 
First we consider the case in which an initially isotropic trap 
($\omega_{\rho}=\omega_z$) is completely removed 
($\omega_{\rho} =\omega_z=0$) for $t>0$. Then, at positive 
time, $V_{ext}^{>}$ only contains the gravitational 
term and the barrier. 
In cylindrical coordinates, the TDGPE equation becomes 
$$ 
i\hbar {\partial \over \partial t} \psi(\rho,z,t) 
= \left[ -{\hbar^2 \over 2 m} \left( 
{\partial^2 \over \partial \rho^2 } 
+ {1 \over \rho} {\partial \over \partial \rho} 
+ {\partial^2 \over \partial z^2} \right) \right. 
$$
\beq 
+ V_{ext}^{>}(\rho,z) + \gamma 
|\psi(\rho,z,t)|^2 \Big] \psi(\rho,z,t) \; ,  
\eeq 
which is a nonlinear 
parabolic partial differential equation in $2+1$ dimensions 
and we study the time evolution 
starting from the ground state of the system with potential (1). 
\par 
The numerical integration of this cylindrical 
TDGPE is performed using a modified split operator technique, 
adapted to the integration of a Schr\"odinger equation [14]. 
We write the Eq. (2) in the form 
\beq 
i \hbar \frac{\partial}{\partial t} \psi(\rho,z, t) =
\left[ \hat{H}_{\rho}(\rho,z,t) + \hat{H}_{z}(\rho,z,t) \right] 
\psi(\rho,z,t) \; , 
\eeq 
where 
\beq 
\hat{H}_{\rho} (\rho,z,t) =  
{\hat T}_{\rho} + V_{ext}^{>}(\rho,z=0) 
+ \frac{2}{3} \gamma |\psi(\rho,z,t)|^{2} \; ,
\eeq 
\beq 
\hat{H}_{z} (\rho,z,t) = 
{\hat T}_z + V_{ext}^{>}(\rho=0,z) 
+ \frac{1}{3} \gamma |\psi(\rho,z,t)| ^{2} \; ,
\eeq 
with $\hat{T}_{\rho}$ and $\hat{T}_z$ the transverse and 
axial kinetic operators, respectively. 
The full Hamiltonian is split in two sub-Hamiltonians, 
so that at each time we have to write the Laplacian and the 
external potential with respect to one coordinate only, 
leading to the solution of a
tridiagonal system, and to huge savings in computer memory. 
Equation (3) is integrated using the scheme 
$$ 
\psi (\rho,z,t+\delta) = 
\left[1+\hat{A}_{z}(t) \right]^{-1} \left[1-\hat{A}_{\rho}(t)\right] 
$$
\beq
\times \left[1+\hat{A}_{\rho}(t) \right]^{-1} 
\left[1-\hat{A}_{z}(t)\right] \psi (\rho,z,t) \; , 
\eeq 
where $\delta$ is the integration time step and 
$\hat{A}_{j}(t)\equiv i \delta \hat{H}_{j}(\rho,z,t)/(2\hbar)$ 
with $j=\rho,z$. The splitting is carried out so that the 
commutators are exact up to the order $\delta ^{2}$ included. 
The nonlinear term $\gamma |\psi (\rho,z,t)|^{2}$ gives a problem, 
because we should really use a $\psi$ somehow averaged over 
the time step $\delta$, not a $\psi$ evaluated at the beginning of the
time step. To circumvent this problem, we used a predictor
corrector method. Each integration step $t\to t+\delta$ 
is performed twice: the first time we 
use $\psi (\rho,z,t)$ in the nonlinear term, obtaining a 
predicted $\tilde \psi (\rho,z,t+\delta)$; we then repeat the 
integration step, starting again from $\psi (\rho,z,t)$, but using 
$\frac{1}{2} \left (\psi (\rho,z,t) + 
\tilde \psi (\rho,z,t+\delta) \right )$ in the nonlinear term. 
In this way the solution is accurate to O$(\delta^2)$. 
At each time step the matrix elements entering 
the Hamiltonian are evaluated by means of 
finite-difference approximants using a typical mesh of 
$400\times 2400$ points in the ($\rho,z$) plane. 
As a check of the accuracy of the algorithm we find that 
in the time interval $\Delta t=10 \omega_H^{-1}$ 
the normalization of $\psi$ is conserved within $5$ $^{0}/_{00}$ 
and the energy within few $\%$. 
Note that Eq. (3), now with the full potential (1), 
can be used with an imaginary time $t=i\tau$ 
to find the ground-state of the system. 
\par 
In our calculations we adopt the harmonic oscillator units 
chosing $\omega_H = (\omega_{\rho}^2\omega_z)^{1/3} 
= 2\pi \times 100$ Hz. For $^{23}$Na atoms, the harmonic length is 
$a_H=(\hbar / (m \omega_H))^{1/2} = 27$ $\mu$m and 
the scattering length is $a_s=3$ nm. The nonlinear term 
in scaled units is given by ${4\pi a_s(N-1)/a_H}$. 

\section{Results and discussion} 

First we study the case in which the initial trap is isotropic 
($\omega_{\rho}=\omega_z =\omega_H$). 
In our computations we set $z_0=15 a_H$ so that 
the condensate is initially far from the Gaussian 
potential barrier whose effect is negligible. 
Actually, also the gravitational effect is negligible 
and the density profile of the ground-state has 
a Gaussian shape for a small number of atoms 
and it is an inverted parabola for a large number 
of particle. Then, we switch off the harmonic potential 
and use the previous wave-function as initial 
condition for Eq. (3). The total energy per particle 
of the condensate is about $180$ $\hbar\omega_H$ with $N$ ranging from 
$1$ to $10^{5}$. 

\begin{figure}
\centerline{\psfig{file=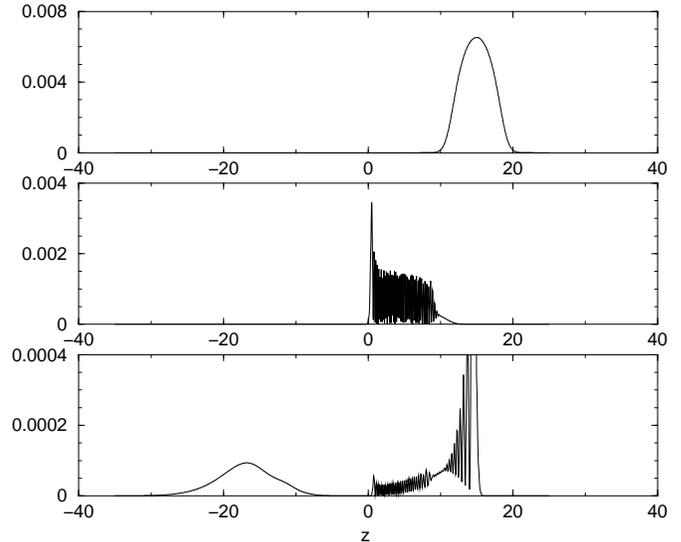,height=2.9in}}
\caption{Density profile $|\psi(\rho=0,z)|^2$ at 
$t=0$, $t=2$ and $t=4$. 
Barrier parameters: $U=200$ and $\sigma=1$. 
$N=10^5$ condensed atoms. 
Lenght in units $a_H=27$ $\mu$m, time in units $\omega_H^{-1}=1.6$ ms and 
energy in units $\hbar \omega_H$.} 
\end{figure} 

\par 
To simulate an impenetrable wall, we have chosen the following 
values for the parameters of the Gaussian potential barrier: 
$U=1000$ $\hbar \omega_H$ and $\sigma = a_H$. 
The center of mass motion of the condensate 
is quasi-periodic due to the bouncing of the condensate. 
Because the condensate is also expanding isotropically, 
the oscillation along the $z$ axis is damped and this damping 
increases with the number of atoms (larger repulsive interaction). 
We have verified that interference fringes appear when 
the condensate collides on the barrier. 
\par 
The phenomenon of interference clearly appears also when 
the energy $U$ of the barrier is comparable to the chemical potential 
$\mu$ of the condensate defined as the 
total energy per particle plus the interaction energy 
per particle. The chemical potential, in contrast 
to the energy, is time-dependent [15]. Nevertheless, 
in our calculations $\mu$ is practically constant, apart 
a slight increase (no more than $4\%$) during the collision time 
due to the enhanced density of the condensate. 
In Fig. 1, we plot the density profile of 
the condensate along the symmetry axis at three different instants. 
In this case the energy barrier is $U=200$ $\hbar \omega_H$ 
and the initial chemical potential $\mu = 184.23$ $\hbar \omega_H$.  
Besides the interference between the incident wave-function 
and the reflected wave-function, one sees a transmitted wave, 
due to the quantum tunneling of the condensate. 
\par 
For a non-interacting gas the semiclassical 
approximation of the linear Schr\"odinger equation 
predicts that the tunneling probability is proportional to 
$e^{-2\int_a^b dz\sqrt{2m(V_{ext}^{>}(z)-\bar{E})/\hbar^2}}$, 
where $a$ and $b$ are the classical turning points and $\bar{E}$ is the 
energy per particle of the condensate that, in the 
non-interacting case, coincides with the chemical potential $\mu$. 
In the interacting case, on the basis of a WKB study 
of the 1D GPE in the weak-coupling regime [8], 
it has been shown that the energy per particle $\bar{E}$ 
should be substituted by the chemical potential $\mu$ 
of the Bose condensate. 
\par 
In general, the semiclassical quantization 
of the nonlinear Schr\"odinger equation 
appears a difficult task but numerical results 
can be easly obtained with our algorithm. As shown in Fig. 2, 
the effect of nonlinear term is such that the tunneling 
fraction $P_T$ increases with the number $N$ of condensed atoms, 
i.e. by increasing the chemical potential. 
For instance, for $N=10^5$ the tunneling probability is 
about $4$ time larger than the non-interacting case. 
Note that, due to the high density in the impact region, 
the self-interaction becomes quite strong 
for a large number of particles and this clearly affects 
the tunneling fraction, as shown in Fig. 2. 
\par 
While the transmitted wave-function of the condensate 
falls towards $z=-\infty$, the reflected wave-function 
goes back close to the initial position $z_0$ and then falls down 
again under the force of gravity. 
It means that, after a time roughly equal to the 
period $2 \sqrt{2z_0/g}$, the condensate scatters 
again on the Gaussian barrier and now the tunneling fraction 
is smaller. 

\begin{figure}
\centerline{\psfig{file=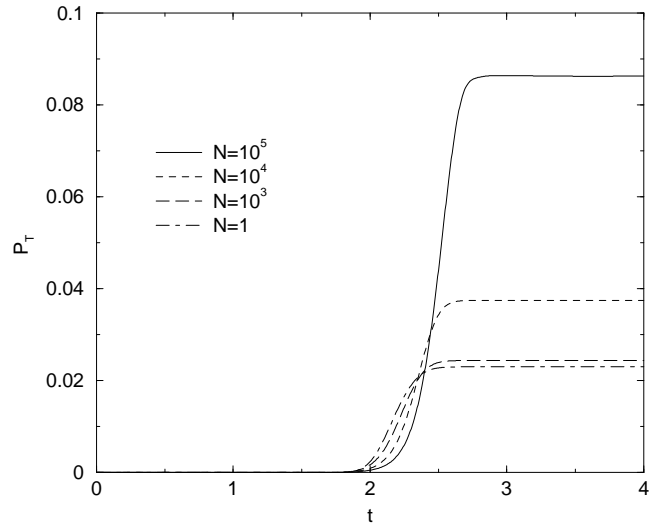,height=2.8in}}
\caption{Tunneling fraction $P_T$ as a function of time $t$. 
Barrier parameters: $U=200$ and $\sigma=1$. 
$N=1$ means non-interacting case. 
Initial position: $z_0=15$. Units as in Fig 1.} 
\end{figure}

In Fig. 3 we plot the tunneling fraction $P_T$ 
as a function of time, showing that the macroscopic quantum tunneling 
of a Bose condensate falling under gravity is a quasi-periodic 
phenomenon. This shows that experimental set up similar 
to what we are considering should be able to generate 
coherent atomic pulses. 

\begin{figure}
\centerline{\psfig{file=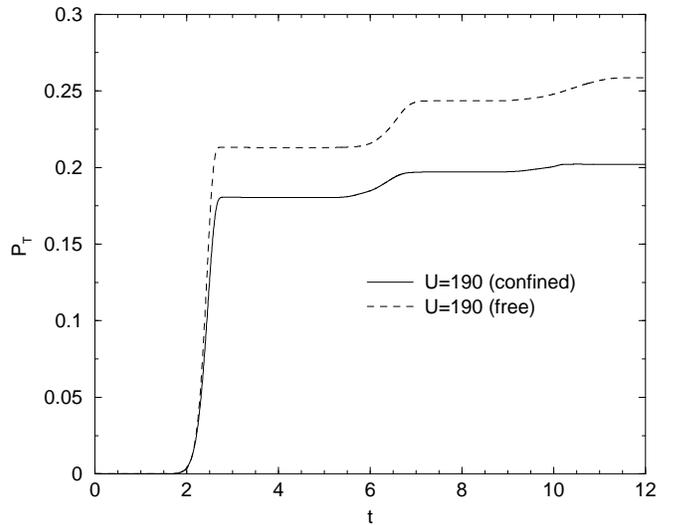,height=2.8in}}
\caption{Tunneling fraction $P_T$ as a function of time $t$. 
Comparison between radially-free ($\omega_{\rho}=0$) 
and under radial confinament ($\omega_{\rho}=0.5$) falling 
condensates. $N=10^5$ condensed atoms and 
initial position $z_0=15$. Barrier width: 
$\sigma =1$. Units as in Fig 1.} 
\end{figure}

\par
In the previous computation the condensate expands for $t>0$ 
in all directions but this leads to a decrease in density 
at the second bounce with a reduced tunneling fraction. 
It is interesting to see what happens 
if the condensate remains trapped in the transverse directions. 
We obtain this assuming that $V_{ext}^{>}$ does not confine 
($\omega_z =0$) in the $z$ direction but it remains confining 
($\omega_{\rho}=\alpha \omega_H$) in the horizontal directions. 

\begin{figure}
\centerline{\psfig{file=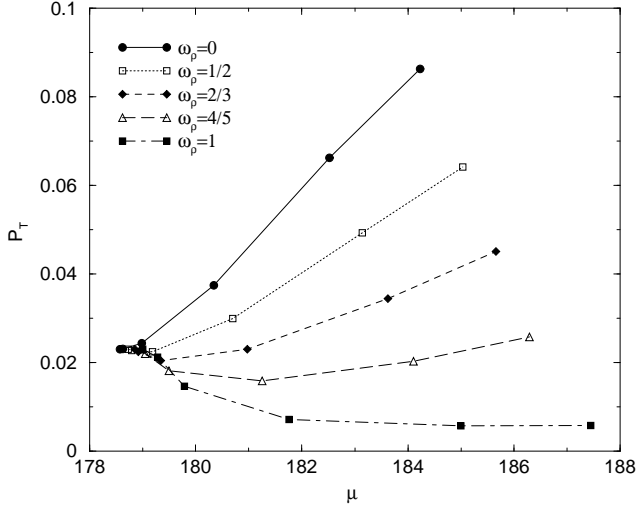,height=2.7in}}
\caption{Tunneling fraction $P_T$ as a function of 
the initial chemical potential $\mu$ for different 
frequencies $\omega_{\rho}$ of radial confinament. 
The points in each line correspond to 
a sequence of numbers $N$ of atoms. From left to right: 
$N=1,10^2,10^3,10^4,5\times 10^4,10^5$. 
Barrier parameters: $U=200$ and $\sigma=1$. 
Units as in Fig 1.} 
\end{figure} 

From Fig. 4 we see that, depending on the degree of 
this radial confinement, the interatomic interaction 
(and the chemical potential $\mu$) either 
enhances or depresses the tunneling probability $P_T$. 
Note that, in the present case, the system is not quasi-1D, 
due to the isotropic initial condition. 
In the case of strong radial confinament, the falling condensate 
shrinks in the radial direction during the motion. 
Such compression modifies the effect of the nonlinear term in the TDGPE. 
As shown in Fig. 4, the tunneling probability $P_T$ initially 
decreases and then increases with the chemical potential $\mu$. 
\par 
In the previous calculations 
we have considered Bose condensates with 
isotropic initial conditions ($\omega_{z}=\omega_{\rho}$). 
Often experimentally the confining potential of the trap 
is anisotropic. Here we investigate the case of 
cigar-shaped initial conditions ($\omega_{z}=\omega_{\rho}/10$) 
which model a quasi-1D system. 
In this case we find that the tunneling 
fraction increases by increasing the number $N$ 
of particles both in the case of absence of confinement 
($\omega_{\rho}=0$) and in the case of strong radial 
confinement (Fig. 5). 
As shown in Fig. 5, due to the anisotropy of the trap, 
the axial elongation of the condensate grows with the 
number $N$ of particles and for $N=10^4$ one tip of the 
condensate touches the Gaussian barrier. 

\begin{figure}
\centerline{\psfig{file=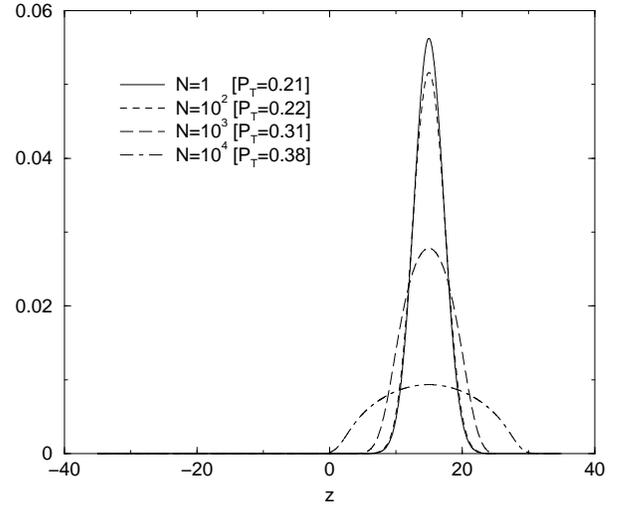,height=2.7in}}
\caption{Density profile $|\psi(\rho=0,z)|^2$ at $t=0$ 
for various values of the number $N$ of condensed atoms.  
Barrier parameters: $U=200$ and $\sigma=1$. 
Radially-confined ($\omega_{\rho}=1$) falling condensate 
with cigar-shaped ($\omega_{\rho}/\omega_{z}=10$) 
initial condition. $P_T$ is the tunneling fraction. 
Units as in Fig 1.} 
\end{figure} 

The exponential law suggested by 
Zapata et al. [8] for a 1D system seems able to capture the behavior 
of $P_T$ as a function of the chemical potential $\mu$: 
on the basis of this law one finds 
a relative increment of the tunneling fraction of $87\%$ 
as $N$ changes from $1$ to $10^4$, whereas this relative increment 
is about $81\%$ on the basis of the TDGPE. 
\par 
In Fig. 6 we plot the density profile of the condensate 
at three time step. One sees that the transmitted pulse has a 
peculiar feature: the density profile clearly shows two peaks 
(see Fig. 6). The amplitude of the second slower peak increases 
with the number of particles. 

\begin{figure}
\centerline{\psfig{file=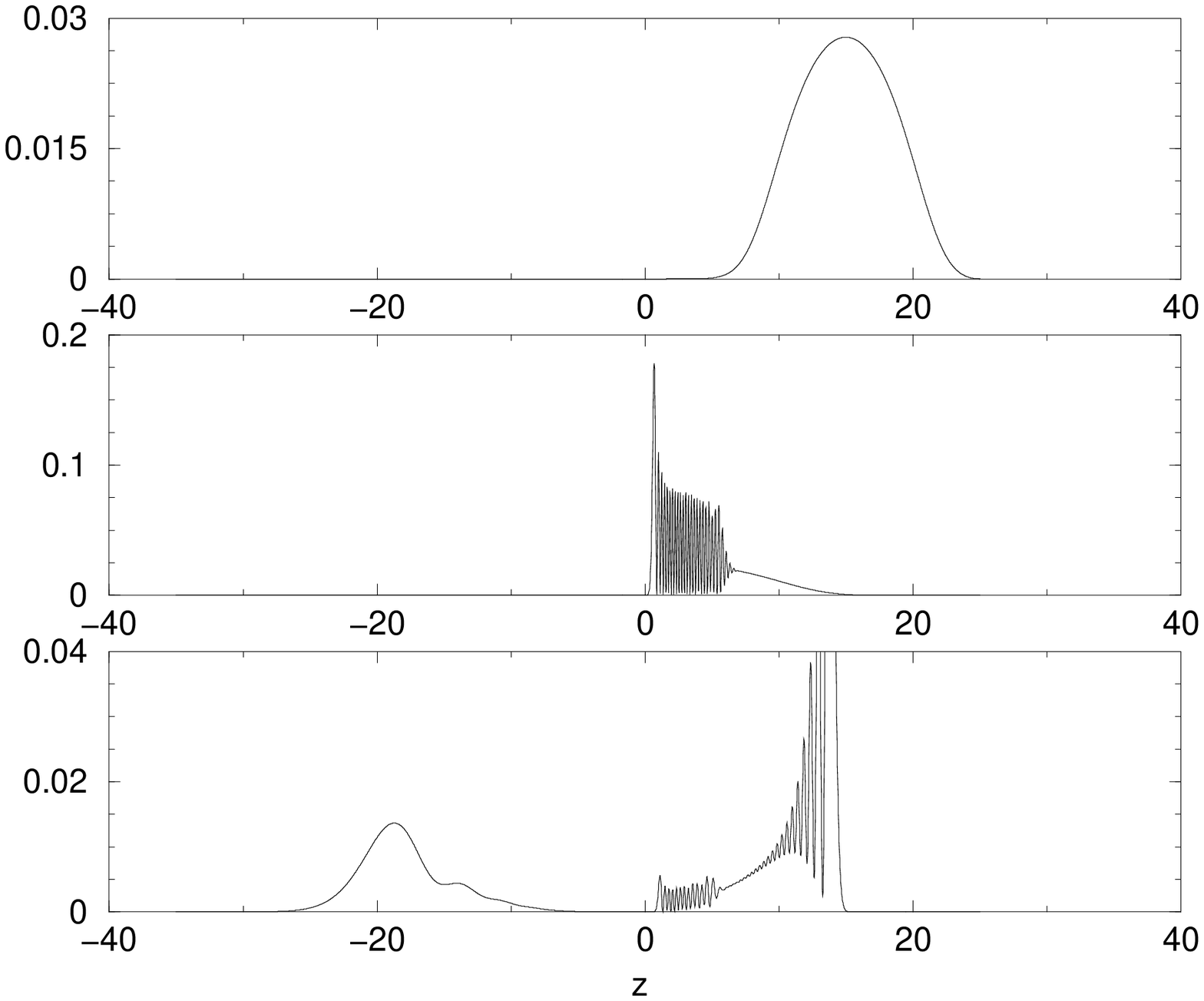,height=2.9in}}
\caption{Density profile $|\psi(\rho=0,z)|^2$ at 
$t=0$, $t=2$ and $t=4$. 
Barrier parameters: $U=200$ and $\sigma=1$. 
$N=10^3$ condensed atoms. 
Radially-confined ($\omega_{\rho}=1$) falling condensate 
with cigar-shaped ($\omega_{\rho}/\omega_{z}=10$) 
initial condition. Units as in Fig 1.} 
\end{figure}

Actually, for a large number 
of particles a third small peak can be detected. 
In the case of an isotropic cloud, the second peak 
shows up only with $N\geq 10^5$ particles (see Fig. 1). 
These two features of the tunneling of a cigar-shaped cloud, 
a tunneling rate enhanced by the interaction and the presence 
of a second peak in the transmitted pulse, are also present 
in the case of a purely one dimensional system. 
\par 
For the sake of completeness, we have also considered 
the single-vortex state $\psi({\bf r},t) = 
\rho \phi(\rho,z) e^{i\theta}$ as initial condition 
but we have not found any 
significant difference in the dynamics of the condensate. 
\par 
We have presented the results for the Na atom. 
As long as the scattering length is positive our 
results apply also to the other alkali-metal atoms by suitably 
rescaling the spatial and temporal scales and the number 
of particles. 

\section{Conclusions} 

We have studied the dynamics of Bose condensates 
falling under gravity and scattering on a Gaussian barrier 
that models a mirror of light. 
Apart the pure bouncing with interference, 
that can be seen for very large values 
of the energy barrier, we have investigated 
quantum tunneling. Our results show that 
macroscopic quantum tunneling of a Bose condensate 
should be observable experimentally and that the tunneling 
rate is very sensitive to the interatomic interaction and 
to geometrical features of the experiment. 
If initially the cloud is spherical under free fall, 
the tunneling fraction grows with the number of particles, 
which increases the interatomic interaction and the 
chemical potential of the condensate. 
But, the fraction is reduced 
by the interatomic interaction if the 
falling condensate is under a strong enough transverse 
confinement. In a 1D case, and in the case of a condensate 
with a cigar-shaped initial condition, 
the interatomic interaction always enhances 
the tunneling rate. Therefore the tunneling rate 
becomes larger and larger as the number of particles increases. 
In addition, in the 1D or quasi 1D case the density profile 
of the transmitted wave shows two peaks. 
We have also shown that in our system 
macroscopic quantum tunneling is a quasi-periodic phenomenon and 
it can be used to generate Bose condensed atomic pulses. 
One could generate truely periodic pulses if the bouncing part 
of the condensate is reconfined for a short period of time 
when it returns in the initial position and additional atoms 
are condensed in order to regenerate the initial state. 
Often the dynamics and static of a condensate 
is modelled by a 1D system. One important result of our 
computation is that geometrical aspects are important in 
tunneling phenomena and that 1D results not 
always reflect the behavior of 3D systems. 

\section*{References}

\begin{description}

\item{\ [1]} M.H. Anderson, J.R. Ensher, M.R. Matthews, C.E. Wieman, 
and E.A. Cornell, Science {\bf 269} 189 (1995).  

\item{\ [2]} K.B. Davis, M.O. Mewes, M.R. Andrews, N.J. 
van Druten, S.D. Drufee, D.M. Kurn, and W. Ketterle, 
Phys. Rev. Lett. {\bf 75} 3969 (1995). 

\item{\ [3]} C.C. Bradley, C.A. Sackett, J.J. Tollett, and 
R.G. Hulet , Phys. Rev. Lett. {\bf 75} 1687 (1995). 

\item{\ [4]} F. Dalfovo, S. Giorgini, L. Pitaevskii, and S. Stringari, 
Rev. Mod. Phys. {\bf 71}, 463 (1999). 

\item{\ [5]} M.R. Andrews, C.G. Townsend, H.J. Miesner, 
D.S. Drufee, D.M. Kurn, and W. Ketterle, Science {\bf 275}, 637 (1997). 

\item{\ [6]} A. Smerzi, S. Fantoni, S. Giovannazzi and 
S.R. Shenoy, Phys. Rev. Lett. {\bf 79}, 4950 (1997). 

\item{\ [7]} G.J. Milburn, J. Corney, E. Wright and 
D.F. Walls, Phys. Rev. A {\bf 55}, 4318 (1997).  

\item{\ [8]} I. Zapata, F. Sols and A.J. Leggett, 
Phys. Rev. A {\bf 57}, R28 (1998). 

\item{\ [9]} S. Raghavan, A. Smerzi, S. Fantoni and S.R. 
Shenoy, Phys. Rev. A {\bf 59}, 620 (1999). 

\item{\ [10]} L. Salasnich, A. Parola, L. Reatto, 
Phys. Rev. A {\bf 60}, 4171 (1999). 

\item{\ [11]} K. Bongs, S. Burger, G. Birkl, K. Sengstock, 
W. Ertmer, K. Rzazewski, A. Sampera, and M. Lewenstein, 
Phys. Rev. Lett. {\bf 83}, 3577 (1999). 

\item{\ [12]} E.P. Gross, Nuovo Cimento 20(1961) 454; 
J. Math. Phys. {\bf 4}, 195 (1963). 

\item{\ [13]} L.P. Pitaevskii, Zh. Eksp. Teor. Fiz. {\bf 40}, 
646 (1961) [Sov. Phys. JETP {\bf 13}, 451 (1961)]. 

\item{\ [14]} E. Cerboneschi, R. Mannella, E. Arimondo, and 
L. Salasnich, Phys. Lett. A {\bf 249}, 245 (1998). 

\item{\ [15]} C.W. Gardiner, Phys. Rev. A {\bf 56}, 1414 (1997). 

\end{description} 

\end{document}